\documentclass[superscriptaddress,showkeys,notitlepage,onecolumn]{revtex4-1}%
\usepackage{amsmath,amssymb,amsbsy,subfigure,hyperref,subfigure,bbm,times,txfonts,accents,float}
\usepackage{graphicx}
\usepackage{dcolumn}
\usepackage{bm}
\emergencystretch=\maxdimen
\hypersetup{colorlinks=true,linkcolor=blue,anchorcolor=blue,citecolor=blue}

\begin{document}

\title{Phase Anti-Synchronization dynamics between Mechanical Oscillator and Atomic Ensemble within a Fabry-Perot Cavity}

\author{Shao-Qiang Ma}
\affiliation{%
 Key Laboratory of Micro-Nano Measurement-Manipulation and Physics (Ministry of Education), School of Physics and Nuclear Energy Engineering, Beihang University, Xueyuan Road No. 37, Beijing 100191, China
}%
\author{Xiao Zheng}
\affiliation{%
 Key Laboratory of Micro-Nano Measurement-Manipulation and Physics (Ministry of Education), School of Physics and Nuclear Energy Engineering, Beihang University, Xueyuan Road No. 37, Beijing 100191, China
}%
\author{Guo-Feng Zhang}%
 \email{gf1978zhang@buaa.edu.cn}
\affiliation{%
 Key Laboratory of Micro-Nano Measurement-Manipulation and Physics (Ministry of Education), School of Physics and Nuclear Energy Engineering, Beihang University, Xueyuan Road No. 37, Beijing 100191, China
}%

\date{\today}

\begin{abstract}
Phase synchronization refers to a kind of collective phenomenon that the phase difference between two or more systems is locked, and it has widely been investigated between systems with the identical physical properties, such as the synchronization between mechanical oscillators and the synchronization between atomic ensembles. Here, we investigate the synchronization behavior between the mechanical oscillator and the atomic ensemble, the systems with different physical properties, and observe a novel synchronization phenomenon, i.e., the phase sum, instead of phase difference, is locked. We refer to this distinct synchronization as the phase anti-synchronization, and show that the phase anti-synchronization can be achieved in both the classical and quantum level, which means this novel collective behavior can be tested in experiment. Also, some interesting connections between phase anti-synchronization and quantum correlation is found.
\end{abstract}
\keywords{phase synchronization, collective phenomenon, mechanical oscillators, anti-synchronization}
\maketitle
\setcounter{secnumdepth}{6}
\renewcommand\thesection{\Roman{section}}%
\section{Introduction}
Spontaneous synchronization, a collective dynamical behavior of highly interest and relevant, refers to the phenomenon that two or more subsystems synchronized their motions only due to the weak and non-liner interaction between them, without any external time-dependent driving force \cite{1,2,3,4}. The synchronization phenomenon was firstly observed by Huygens in seventeenth century in two coupled pendulum clocks \cite{1}, and after that, this phenomenon has been found in many different physical settings, ranging from sociology and biology to physics. For instance, the collective lightning of fireflies, the beating of heart cells and chemical reaction \cite{5,6,7,8}.

Since the first observation of the occurrence of synchronization, lots of works have been done to investigate this phenomenon \cite{9,10,11,12,13,14,15,16,17}. In classical level, the dynamics of synchronization has been well studied and a standard theoretical system has been constructed to judge whether two classical systems are synchronized \cite{9,10,11,12}. However, the conclusions deduced in classical level as well as the related concept, due to the existence of uncertainty relations, cannot be extended to the quantum level \cite{13,14}. Thus, there still exist some difficulty in the investigation of the quantum synchronization. Nevertheless, a rapid development of quantum synchronization, in recent, has been made \cite{18,19,20,21,22,23}. Complete synchronization and phase synchronization are two common classical synchronization forms. According to \cite{13},  the complete synchronization is achieved when two subsystems acquire identical trajectories under the effects of mutual interactions, and the phase synchronization is instead achieved only the phase difference is locked. In Ref.\cite{13},  Mari, et. al. constructed two quantum measures of these two synchronization forms. They showed that the Heisenberg principle \cite{24,25} set a universal bound to complete synchronization and the measure of phase synchronization is, in principle, unbounded with the help of squeezing resource. The results have been confirmed by Ref. \cite{14} in optomechanical system in numerical way. Meanwhile, the physics essence of the classical phase synchronization has been provided by Ref.\cite{15} from the perspective of quantum mechanics, and they show that the occurrence of classical phase synchronization can be explained as that a mode in the single leaking picture is driven to the ground state. Also, the relationship between the quantum synchronization and quantum correlation has been investigated \cite{18}. All of these powerful works greatly promote the development of the quantum synchronization.

However, the previous investigations are mainly focused on the synchronization between physical systems with the  identical physical properties, for instance, the synchronization between two mechanical oscillators in Refs. \cite{13,14,15,16}, and the synchronization between two atomic ensembles in Refs. \cite{17,26,27}. Few works have been done to investigate the synchronization between the system with different  physical properties. Thus, we wonder that can the synchronization phenomenon occur between different physical platforms, and if so, do there exist some novel and distinct behaviors, which cannot be observed in the previous synchronization phenomenon between the identical physical systems. Based on these motivations, we investigate the synchronization dynamics between mechanical oscillator and atomic ensemble.

The phase synchronization refers to the phenomenon that the phase difference between two synchronized systems is locked. There exist two necessary conditions for the occurrence of the phase synchronization: (i) the interaction between the systems of interest should be non-linear, and (ii) the frequency difference between the systems of interest should be small. To satisfy the two conditions above, we choose the hybrid optomechanical system, which is formed by an atomic gas and a mechanical oscillator placed in a common cavity, as the platform to investigate the synchronization between mechanical oscillator and atomic ensemble, as shown in Fig.\ref{f1}.

In the optmechanics system \cite{28,29,30,31,3a1,3a2,3a3}, the radiation pressure coupling between the optics mode and the mechanics mode is nonlinear \cite{13,14}. The mechanical oscillator and atomic ensemble are indirectly coupled with each other via their interaction with the common intracavity field. The coupling is nonlinear, because, as mentioned above, the coupling between the cavity mode and the mechanics mode is nonlinear. Meanwhile, under the bosonic representation approximation, the atomic ensemble can be considered as a bosonic mode, the frequency of which can be very close to the frequency of the mechanics oscillator \cite{29,30}. Thus, the hybrid optomechanical system satisfies the two necessary conditions for the occurrence of synchronization, and it therefore is a promising platform to investigate the synchronization between mechanical oscillator and atomic ensemble. In such a system, we find a distinct synchronization phenomenon, i.e., the phase sum, instead of the phase difference, of the mechanical oscillator and atomic ensemble is locked after a long-time evolution. We refer to this novel phenomenon as the phase anti-synchronization, and the phase anti-synchronization can be tested experimentally, because the parameter regime of the hybrid optomechanical system is experimental available.

The outline of the paper is as follows. In Sec.\ref{C2}, the physical model is introduced. Also, the dynamics of the system has been investigated. Sec.\ref{C3} is mainly used to investigate the phase anti-synchronization in both the classical and quantum level. In Sec.\ref{C4}, a discussion of the relationship between phase anti-synchronization and quantum correlation is presented. Finally, Sec.\ref{C5} is devoted to the conclusion.

\section{Physical Model} \label{C2}
\begin{figure}
\centering 
\includegraphics[height=7.5cm]{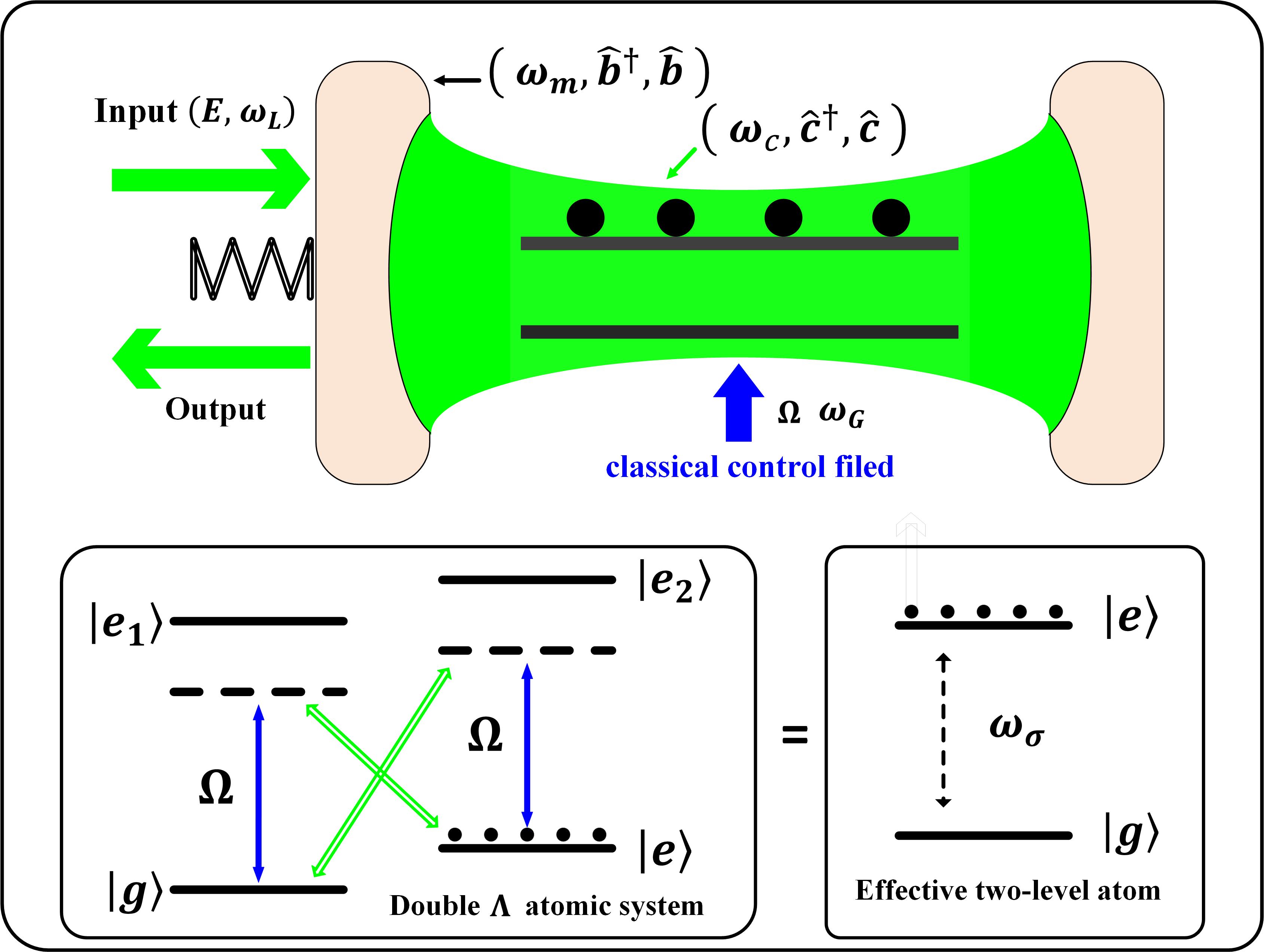}
\caption{Schematic illustration of the hybrid optomechanical system consisting of an ensemble of effective two-level atoms and a mechanical oscillator $(\omega_{m}, \hat{b}^{\dagger}, \hat{b})$ within a common cavity $(\omega_{c}, \hat{c}^{\dagger}, \hat{c})$, which is driven by a laser at frequency $\omega_L$ and amplitude $\eta$. The effective two-level atom is simulated by a coupled double-$\Lambda$ system, which interacts non-resonantly with the cavity mode and is driven by a classical control field with the Rabi frequency $\Omega$ and frequency $\omega_G$. Due to the sufficiently off-resonant interaction, the excited states $\left|e_{1}\right\rangle$ and $\left|e_{2}\right\rangle$ is weakly populated and thus can be adiabatically eliminated. Then, the coupled double-$\Lambda$ system is reduced to an effective two-level system, with the levels $|e\rangle$ and $|g\rangle$. The ensemble of effective two-level atoms is indirectly coupled to the mechanical oscillator via the common cavity.} 
\label{f1}
\end{figure}

As shown in Fig.\ref{f1}, the Fabry-Perot cavity consists of a fixed and a movable mirror, and is driven by a classical laser with frequency $\omega_L$ and amplitude $\eta$. The movable mirror can be considered as a mechanical oscillator. Assume that $\omega_c$ and $\omega_m$ represent the frequencies of the cavity mode and the mechanical mode, and the corresponding annihilation (creation) operators are denoted by $\hat{c}$ $(\hat{c}^\dag)$ and $\hat{b}$ $(\hat{b}^\dag)$, respectively. Then the Hamiltonian of the Fabry-Perot cavity can be written as \cite{29,30}:
\begin{align}\label{1}
\hat{H}=\hat{H}_{c}+\hat{H}_{m}+\hat{H}_{o m}+\hat{H}_{d} \tag{1},
\end{align}
where $\hat{H}_{c}=\hbar \omega_{c} \hat{c}^{\dagger} \hat{c}$ and $\hat{H}_{m}=\hbar \omega_{m} \hat{b}^{\dagger} \hat{b}$ are the free Hamiltonians of the cavity and mechanical modes, respectively. $\hat{H}_{o m}=\hbar g_{m} \hat{c}^{\dagger} \hat{c}(\hat{b}^{\dagger}+\hat{b}) / \sqrt{2}$ describes the optomechanical interaction with $g_m$ being the coupling strength. The last term $\hat{H}_{d}=i \hbar \eta(\hat{c}^{\dagger} e^{-i \omega_{L} t}-\hat{c} e^{i \omega_{L} t})$ is used to express the inputting driving by the classical laser.

In addition to the mechanical oscillator, there also exists an ensemble of $N$ ultracold atoms within the Fabry-Perot cavity. The atomic ensemble interacts non-resonantly with the cavity mode and is driven by a classical control field with the Rabi frequency $\Omega$ and frequency $\omega_G$. As shown in Fig.\ref{f1}, the excited states $\left|e_{1}\right\rangle$ and $\left|e_{2}\right\rangle$ is weakly populated when the interaction is sufficiently off-resonant. Then, the excited states $\left|e_{1}\right\rangle$ and $\left|e_{2}\right\rangle$ can be adiabatically eliminated, and the coupled double-$\Lambda$ system is therefore reduced to an effective two-level system, with two levels $|e\rangle$ and $|g\rangle$. Also, the Zeeman splitting frequency between the two energy levels $|e\rangle$ and $|g\rangle$, denoted by $\omega_\sigma$, can be tuned to be very close to the frequency $\omega_m$ of the mechanical mode by a static external magnetic field \cite{32}. Then, the Hamiltonian of the atomic ensemble can be written as:

\begin{align}\label{2}
\hat{H}_{a t}=\hbar \omega_{\sigma} \hat{S}_{z}+\hbar G_{0} \cos \left(\omega_{G} t\right)\left(\hat{c}^{\dagger}+\hat{c}\right)\left(\hat{S}_{+}+\hat{S}_{-}\right)\tag{2},
\end{align}
where $G_0$ is the atom-cavity coupling strength, $\hat{S}_{z}=\sum_{i=1}^{N}\left(|e\rangle_{i}\left\langle\left. e\right|_{i}-| g\right\rangle_{i}\left\langle\left. g\right|_{i}\right) / 2\right.$, and $\hat{S}_{+}=\sum_{i=1}^{N}|e\rangle_{i}\langle g|_{i}=(\hat{S}_{-})^{\dagger}$ with $|\ldots\rangle_{i}$ being the states of $i$-atom. Assume that the atoms are initially pumped in the hyperfine higher energy level, namely $|e\rangle$ , which leads to an inverted ensemble. The inverted ensemble can be well approximated by a harmonic oscillator when $N$ is large. Then, Hamiltonian of the atomic ensemble can be rewritten as (for more detail,please see the Appendix A):

\begin{align}\label{6}
\hat{H}_{a t}=-\hbar \omega_{\sigma} \hat{d}^{\dagger} \hat{d}+\hbar \frac{G}{2}\left(\hat{c}^{\dagger}+\hat{c}\right)\left(\hat{d}^{\dagger}+\hat{d}\right)\tag{3}.
\end{align}
where $\hat{d}$ is effective atomic bosonic annihilation operator (for more detail,please see the Appendix A), and  $G=G_{0} \sqrt{N}$. Combining the Hamiltonians of the Fabry-Perot cavity and the atomic ensemble, one can obtain the Hamiltonian of the whole system (in the interaction picture with respect to $\hbar \omega_{L} \hat{c}^{\dagger} \hat{c}$):
\begin{align}\label{7}
\hat{H}=&\hbar \Delta \hat{c}^{\dagger} \hat{c}+\hbar \omega_{m} \hat{b}^{\dagger} \hat{b}-\hbar \omega_{\sigma} \hat{d}^{\dagger} \hat{d}+\hbar g_{m} \hat{c}^{\dagger} \hat{c} \hat{q}_{m}\nonumber \\
&+\hbar g_{d}\left(\hat{c}^{\dagger}+\hat{c}\right) \hat{q}_{d}+i \hbar \eta\left(\hat{c}^{\dagger}-\hat{c}\right)\tag{4},
\end{align}
where $g_{d}=G / \sqrt{2}$, and $\Delta=\omega_{c}-\omega_{L}$ represents the detuning between the driving laser and the cavity field. $\hat{q}_{d}=(\hat{d}^{\dagger}+\hat{d}) / \sqrt{2}$, $\hat{p}_{d}=i(\hat{d}^{\dagger}-\hat{d}) / \sqrt{2}$, $\hat{q}_{m}=(\hat{b}^{\dagger}+\hat{b}) / \sqrt{2}$, and $\hat{p}_{m}=i(\hat{b}^{\dagger}-\hat{b}) / \sqrt{2}$ are the dimensionless position and momentum operators satisfying $\left[\hat{q}_{d}, \hat{p}_{d}\right]=\left[\hat{q}_{m}, \hat{p}_{m}\right]=i$.

Taking advantage of the Hamiltonian (\ref{7}) and considering the damping and noise terms, one can obtain the non-linear Heisenberg-Langevin (HL) equation \cite{33}:
\begin{align}\label{8}
&\dot{\hat{c}}=-\Delta i \hat{c}-g_{m} i \hat{c} \hat{q}_{m}-g_{d} i \hat{q}_{d}-\kappa \hat{c}+E+\sqrt{2 \kappa} \hat{c}^{i n};\nonumber \\
&\dot{\hat{p}}_{m}=-\omega_{m} \hat{q}_{m}-g_{m} \hat{c}^{\dagger} \hat{c}-\gamma \hat{p}_{m}+\hat{\xi}; \nonumber \\
&\dot{\hat{p}}_{d}=\omega_{\sigma} \hat{q}_{d}-g_{d}\left(\hat{c}^{\dagger}+\hat{c}\right)-\Gamma \hat{p}_{d}+\sqrt{2 \Gamma} \hat{p}_{d}^{i n};\nonumber \\
&\dot{\hat{q}}_{m}=\omega_{m} \hat{p}_{m};\nonumber \\
&\dot{\hat{q}}_{d}=-\omega_{\sigma} \hat{p}_{d}-\Gamma \hat{q}_{d}+\sqrt{2 \Gamma} \hat{q}_{d}^{i n}\tag{5},
\end{align}
where $\kappa$ is the decay rate of the cavity mode, $\gamma$ is the damping rate of the mechanics mode and $\Gamma$ is the atomic ensemble dephasing rate.  $\hat{c}^{i n}$ denotes the vacuum optical input noise with zero mean value, $\hat{\xi}$  represents the thermal noise expressing the Brownian stochastic force acting on the mechanics mode, $\hat{q}_{d}^{i n}=(\hat{d}^{i n}+\hat{d}^{i n \dagger}) / \sqrt{2}$, and $\hat{p}_{d}^{i n}=i(\hat{d}^{i n \dagger}-\hat{d}^{i n}) / \sqrt{2}$ with $\hat{d}^{i n}$ being the bosonic operator describing the optical vacuum fluctuations affecting the atomic transition \cite{29,31}. The corresponding non-zero correlation functions reads \cite{33,34,35,36,37,38,39,40}:
\begin{align}\label{9}
\langle\hat{c}^{i n}(t) \hat{c}^{i n \dagger}(t^{\prime})\rangle&=\delta(t-t^{\prime});\nonumber \\
\langle\hat{\xi}(t) \hat{\xi}(t^{\prime})+\hat{\xi}(t^{\prime}) \hat{\xi}(t)\rangle&= 2 \gamma(2 \overline{n}+1) \delta(t-t^{\prime}); \nonumber \\
\langle\hat{q}_{d}^{i n}(t) \hat{q}_{d}^{i n}(t^{\prime})\rangle&=\frac{1}{2} \delta(t-t^{\prime});\nonumber \\
\langle\hat{p}_{d}^{i n}(t) \hat{p}_{d}^{i n}(t^{\prime})\rangle&=\frac{1}{2} \delta(t-t^{\prime})\tag{6},
\end{align}
where $\overline{n}=[\exp \left(\hbar \omega_{m} / k_{B} T\right)]^{-1}$ is the expected number of thermal phonons of the mechanics mode at temperature $T$ with $k_B$ being the Boltzmann constant. The dynamics of the system under consideration can fully described by the HL equation (\ref{8}) and the correlation functions (\ref{9}).

\section{Phase anti-Synchronization} \label{C3}
This section is used to investigate the synchronization behavior between the mechanics and the atomic ensemble from the perspective of both the classical and the quantum level. An arbitrary quantum operator $\hat{O}$ can be reformed as $\hat{O}=O+\delta \hat{O}$ by the mean-field approximation \cite{18}, with $O$ being the expectation of $\hat{O}$ . The classical synchronization can be described by the mean field $O$, and the influence of the quantum fluctuation on the classical behavior can be described by the fluctuation operator $\delta \hat{O}$.
\subsection{Classical Phase anti-Synchronization}
Averaging the HL equation (\ref{8}) over the quantum fluctuations, one can deduce the following non-linear differential equations, which is used to describe the dynamics of the mean field \cite{29}:
\begin{align}\label{10}
&\dot{q}_{c}=-\kappa q_{c}+\Delta p_{c}+g_{m} q_{m} p_{c}+\sqrt{2} E;\nonumber \\
&\dot{p}_{c}=-\Delta q-g_{m} q_{m} q_{c}-\sqrt{2} g_{d} q_{d}-\kappa p_{c}; \nonumber \\
&\dot{p}_{m}=-\omega_{m} q_{m}-\frac{1}{2} g_{m}\left(p_{c}^{2}+q_{c}^{2}\right)-\gamma p_{m};\nonumber \\
&\dot{p}_{d}=\omega_{\sigma} q_{d}-\sqrt{2} g_{d} q_{c}-\Gamma p_{d};\nonumber \\
&\dot{q}_{m}=\omega_{m} p_{m}\nonumber; \\
&\dot{q}_{d}=-\omega_{\sigma} p_{d}-\Gamma q_{d}\tag{7},
\end{align}
where $\hat{q}_{c}=(\hat{c}^{\dagger}+\hat{c}) / 2$ and $\hat{p}_{c}=i(\hat{c}^{\dagger}-\hat{c}) / 2$ are the cavity mode quadrature.As mentioned above, the frequency of the atomic ensemble can be tuned to be very close to the frequency of the mechanic mode by a static external magnetic field. The small frequency difference and the non-linear interaction between the atomic ensemble and mechanical oscillator satisfy the necessary condition for the occurrence of the synchronization.

\begin{figure}
\centering 
\includegraphics[height=6cm]{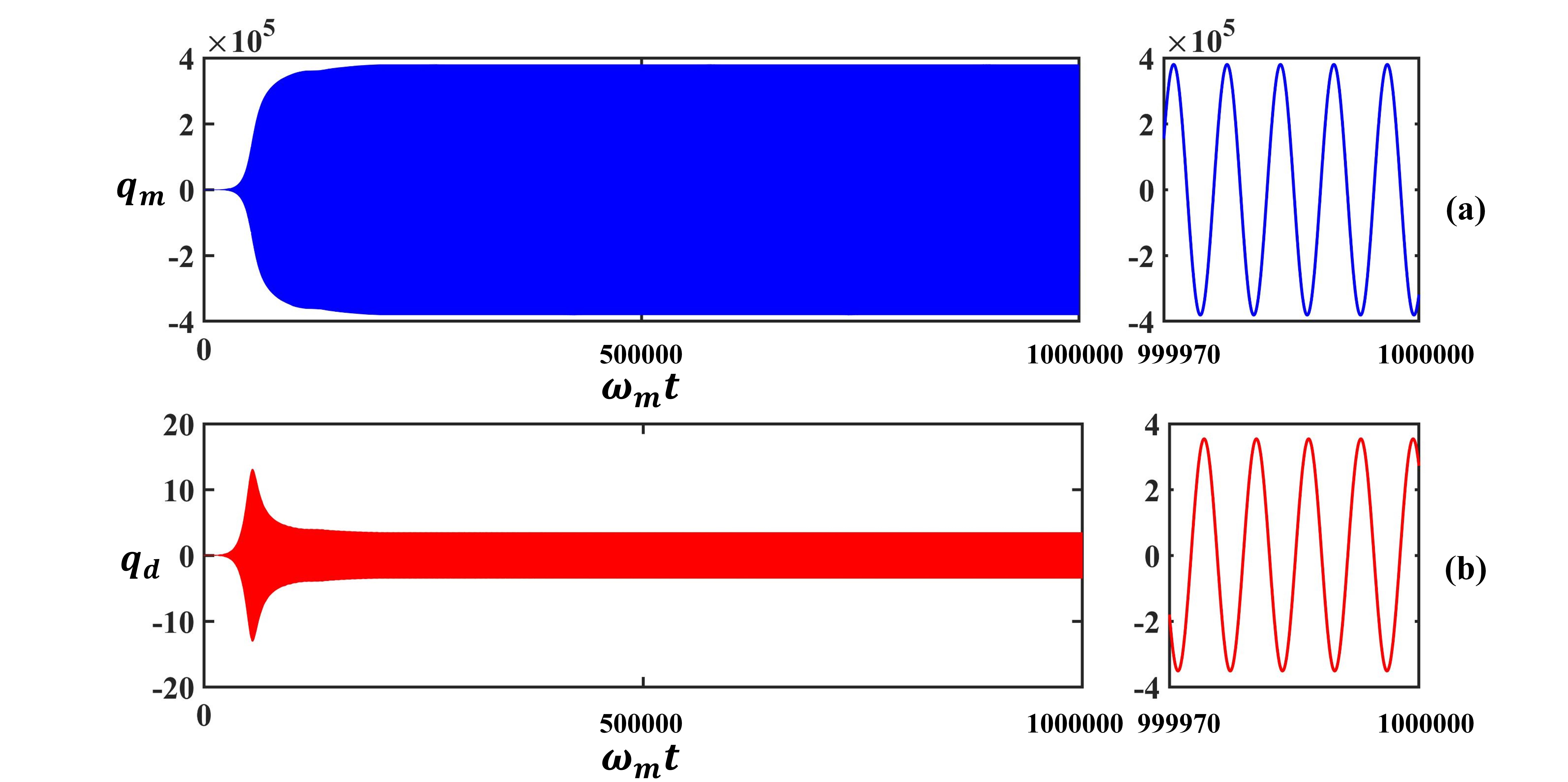}
\caption{Evolutions of $q_m$ and $q_d$ with respective to the scale time $\omega_m t$. Here, the amplitude of the driven laser is taken as $\eta/\omega_m =3000$, and the other related parameters are presented in the main text. } 
\label{f2}
\end{figure}

 The differential equation (\ref{8}) cannot be solved in analytical way, and thus we investigate the phase synchronization in numerical way. The related parameters of the hybrid optomechanical system are taken as $\kappa / \omega_{m}=1,$ $\gamma / \omega_{m}=\Gamma / \omega_{m}=5 \times 10^{-6}$, $\Delta / \omega_{m}=-1$ and $g_{m} / \omega_{m}=g_{d} / \omega_{m}=10^{-5}$, and all of these parameters are experimental evaluable \cite{29,3a1,3a2,3a3}. In such a parameter region, we obtain the time evolutions of the $q_m$ and $q_d$, as shown in Fig.\ref{f2} (a) and (b). One can see that both the atomic ensemble and the mechanical oscillator reach a periodic steady state with a constant amplitude after a period of evolution, and the periodic steady state actually indicates the trajectories of the two subsystem is a closed circle in the phase space, which corresponds to the self-sustained oscillation. All of these provide an ideal platform for the occurrence of phase synchronization.

\begin{figure}
\centering 
\includegraphics[height=7.5cm]{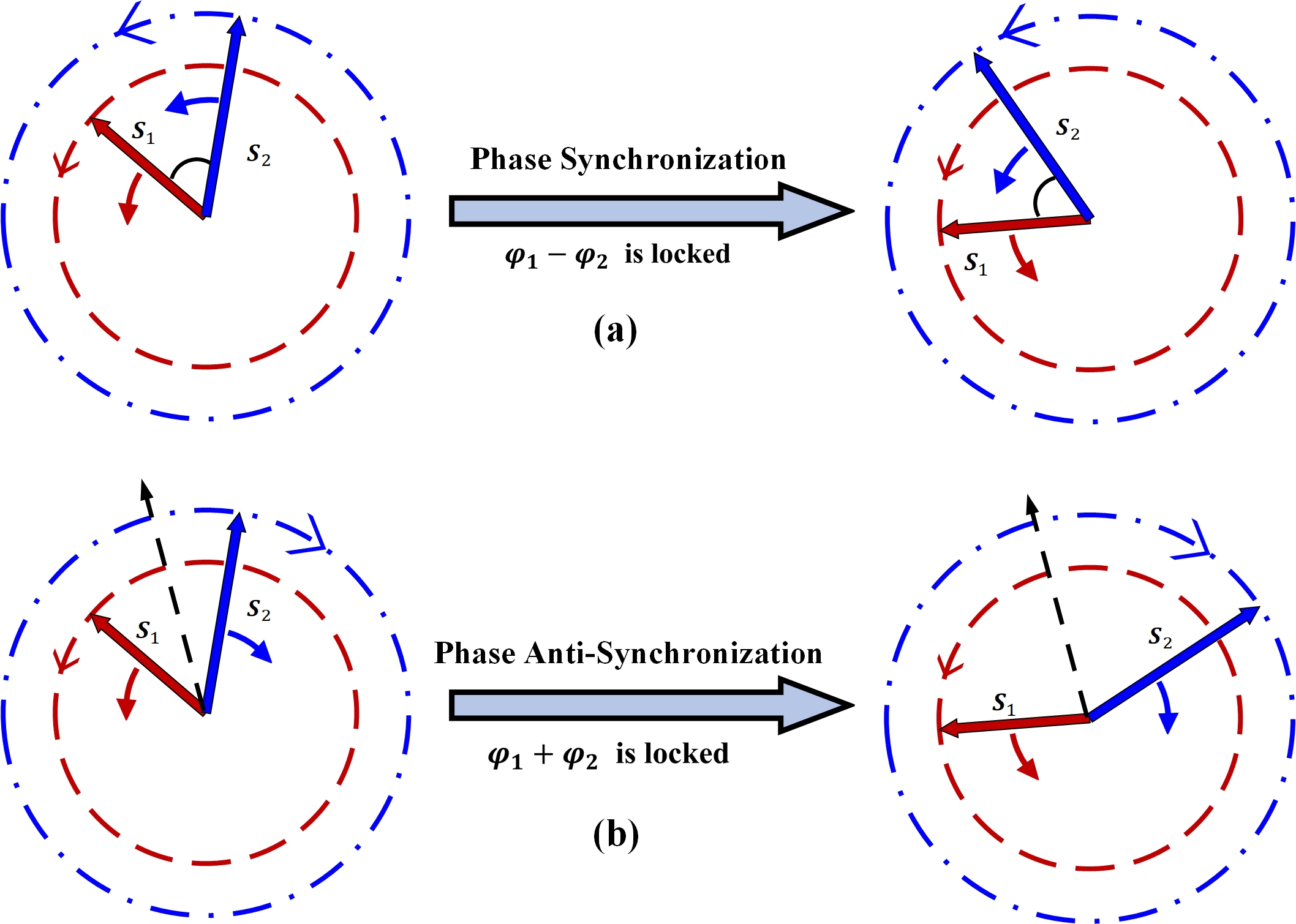}
\caption{The demonstration of traces of the phase synchronization (in (a)) and phase anti-synchronization (in (b)) in the phase space $(q, p)$. The phase synchronization indicates that the difference between $\varphi_1$ and $\varphi_2$, namely $\varphi_1-\varphi_2$, is locked, as shown in (a). Different from phase synchronization, the phase anti-synchronization means that the sum of $\varphi_1$ and $\varphi_2$, namely $\varphi_1+\varphi_2$, is locked, as shown in (b). Here, $\varphi_{j}=\arctan (p_{j} / q_{j})$ represents the phase of the continuous variable subsystem $S_j$ with  $q_j$  and $p_j$  being the canonical variables of $S_j$ } 
\label{f3}
\end{figure}

We then show that the synchronization between the atomic ensemble and mechanical oscillator actually occurs, but the synchronization between them is completely different from the synchronization between the systems with the identical physical properties. The difference between them is demonstrated in Fig.\ref{f3}. Assume two continuous variable subsystems $S_1$ and $S_2$, and the phase of the subsystem $S_j$ takes the form $\varphi_{j}=\arctan (p_{j} / q_{j})$ with $q_j$  and $p_j$  being the canonical variables of $S_j$. The phase synchronization between $S_1$ and $S_2$ occurs when the difference between $\varphi_1$ and $\varphi_2$, namely $\varphi_1-\varphi_2$, is locked, as shown in Fig.\ref{f3} (a). Different from Fig.\ref{f3} (a), Fig.\ref{f3} (b) describes the phenomenon, in which the sum of the phases $\varphi_1+\varphi_2$, instead of the difference of the phases $\varphi_1-\varphi_2$, is locked, and we refer to the synchronization phenomenon illustrated in Fig.\ref{f3} (b) as the phase anti-synchronization. The phase synchronization has been widely investigated and observed in various non-linear systems, but the phase anti-synchronization has merely been observed. Then, we should ask that ``does the phase anti-synchronization exist?".

\begin{figure}
\centering 
\includegraphics[height=6cm]{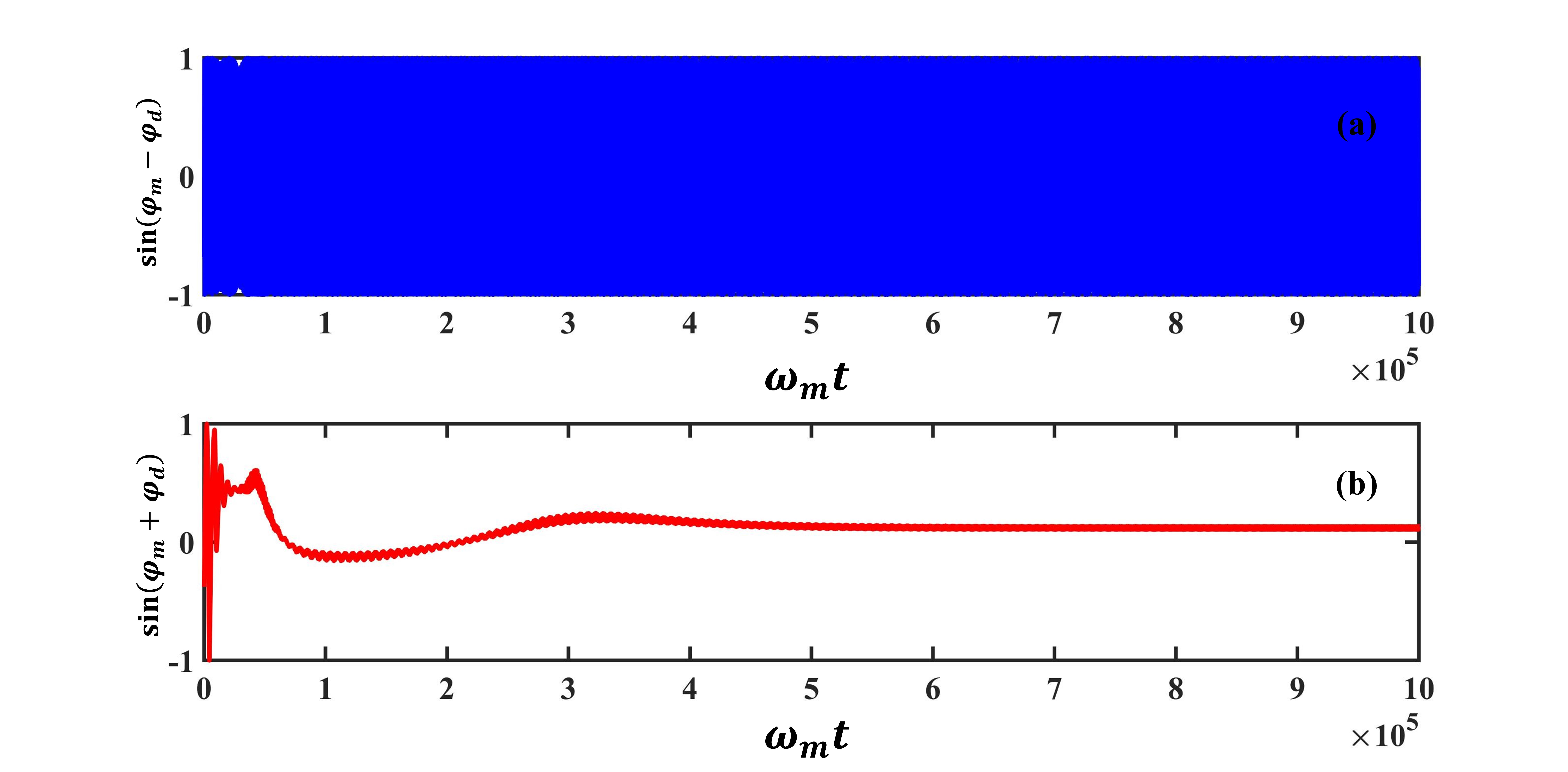}
\caption{Evolutions of $\sin \left(\varphi_{m}-\varphi_{d}\right)$ and $\sin \left(\varphi_{m}+\varphi_{d}\right)$ with respective to the scale time $\omega_m t$. Here, the amplitude of the driven laser is taken as $E/\omega_m =3000$, and the other related parameters are presented in the main text.} 
\label{f4}
\end{figure}

Then, we  show that the synchronization behavior between the atomic ensemble and the mechanic oscillator is the phase anti-synchronization. Here we used $\varphi_{m}$ and $\varphi_{d}$ to represent the phases of  the mechanic oscillator and the atomic ensemble, respectively. The time evolutions of $\sin \left(\varphi_{m}-\varphi_{d}\right)$ and $\sin \left(\varphi_{m}+\varphi_{d}\right)$ is shown in Fig.\ref{f4}. We can see that the sum of the phase is locked, i.e., $\varphi_{m}+\varphi_{d}\cong0.025$, and thus phase anti-synchronization actually occurs between the atomic ensemble and the mechanic oscillator.
\subsection{Quantum  Phase anti-Synchronization}
In this subsection, we mainly investigate the robustness of phase anti-synchronization to the quantum fluctuation and thermal noise, so as to show that the phase anti-synchronization is completely possible in quantum level. For the hybrid optomechanical system under consideration, the influence of the quantum fluctuation on the phase anti-synchronization can be measured by the variance of $\hat{\varphi}_{m}+\hat{\varphi}_{d}$, namely  $\Delta\left(\hat{\varphi}_{m}+\hat{\varphi}_{d}\right)^{2}=\langle\left(\delta \hat{\varphi}_{m}+\delta \hat{\varphi}_{d}\right)^{2}\rangle$ with $\hat{\varphi}_{m}$ and $\hat{\varphi}_{d}$ being the phase operators of the mechanical oscillator and atomic ensemble, respectively. $\delta \hat{\varphi}_{m}$ and $\delta \hat{\varphi}_{d}$ represents the corresponding phase fluctuation operators \cite{14}:
\begin{align}\label{11}
&\delta \hat{\varphi}_{m}=\frac{-\sin \varphi_{m} \delta \hat{q}_{m}+\cos \varphi_{m} \delta \hat{p}_{m}}{\sqrt{2 n_{m}(t)}},\nonumber \\
&\delta \hat{\varphi}_{d}=\frac{-\sin \varphi_{d} \delta \hat{q}_{d}+\cos \varphi_{d} \delta \hat{p}_{d}}{\sqrt{2 n_{d}(t)}}\tag{8},
\end{align}
where $\sqrt{n_{m}(t)}$ and $\sqrt{n_{d}(t)}$ are the amplitude of $q_m$ and  $q_d$, respectively.

Assume that the mechanical oscillator is initially prepared in a thermal state corresponding to the temperature $T$, the atomic ensemble and the cavity mode fluctuations are initially in the vacuum state. Then, one can obtain the numerical result of the $\Delta\left(\hat{\varphi}_{m}+\hat{\varphi}_{d}\right)^{2}$ (for more detail, please see the Appendix B) \cite{37,41,42}, as shown in Fig.\ref{f5}. Fig.\ref{f5} indicates that the variance can be maintained at a stable and small value after a long-time evolution for both $T=0 K$ and $T=10^{-2} K$. That is to say, the classical phase anti-synchronization will not be destroyed by the quantum fluctuation and the thermal noise in low temperature environment. Also, from Fig.\ref{f5}, we can see that the variance at zero temperature is less than the variance at nonzero temperature, and the reason for such a phenomenon is that  the thermal fluctuations is detrimental to the synchronization.
\begin{figure}
\centering 
\includegraphics[height=6cm]{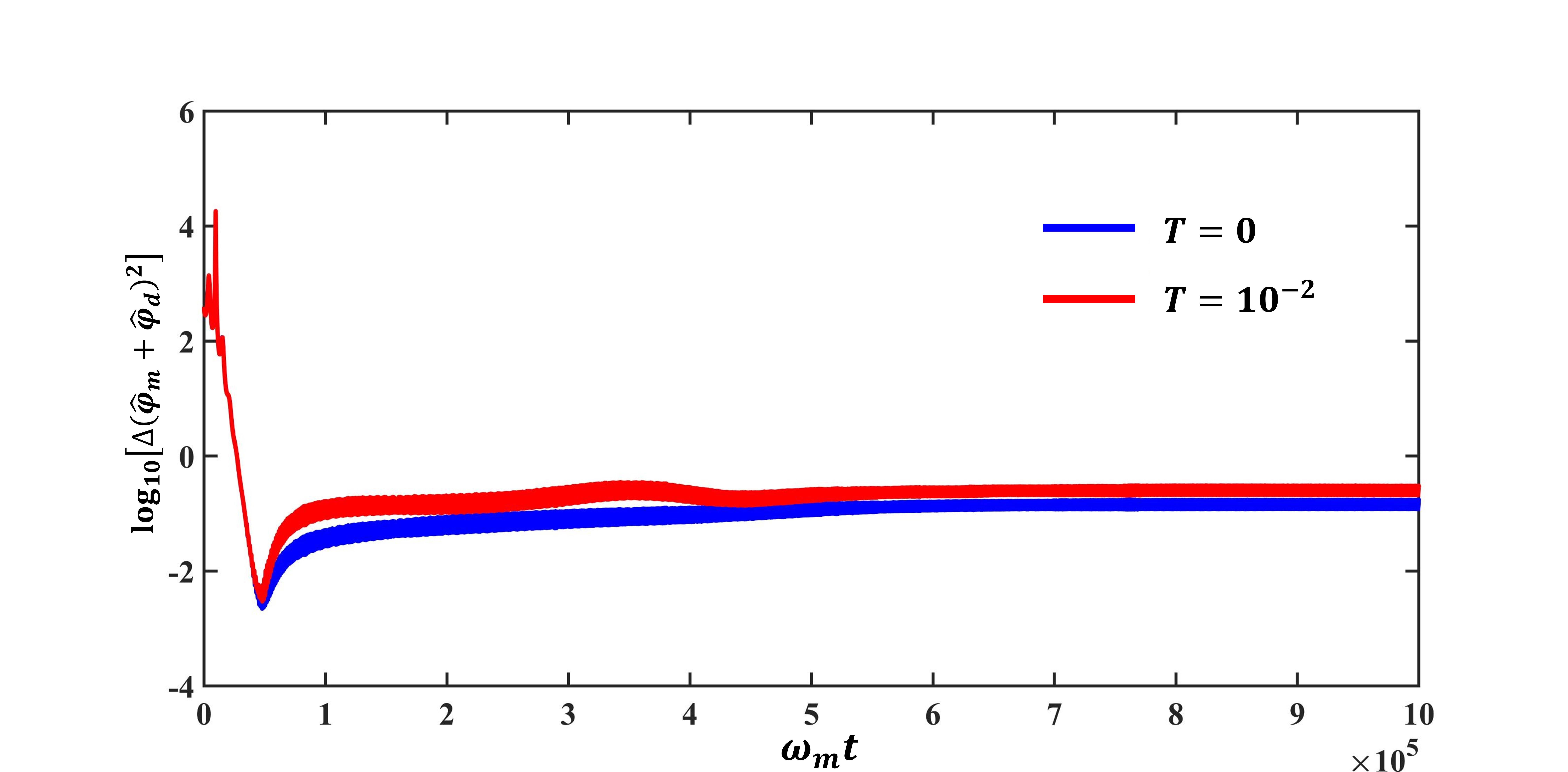}
\caption{Evolutions of $\log _{10}[\Delta(\hat{\varphi}_{m}+\hat{\varphi}_{d})^{2}]$ with respective to the scale time $\omega_m t$ for $T=0 K$ (denoted by blue line) and $T=10^{-2}  K$ (denoted by red line). Here, the amplitude of the driven laser is taken as $\eta/\omega_m =3000$, the frequency is taken as $\omega_m=10^7$ HZ and the other related parameters are presented in the main text.} 
\label{f5}
\end{figure}

 By taking the hybrid optomechanical system as a concrete example, the discussion above shows that the phase anti-synchronization can be observed in such a system. In the following, we will show that the phase anti-synchronization is unbounded in quantum level, from the perspective of quantum squeezing theory. Ref.\cite{13} provided a quantum measure for the phase synchronization, and the measure they provided indicates that the phase synchronization is, in principle, unbounded. Here, we will investigate the relationship between the phase anti-synchronization and the squeezing resource along the way of Ref. \cite{13}. According to Ref.\cite{13}, when the amplitude of the two synchronized systems $S_1$ and $S_2$ are the same \cite{14}, the quantum measure for the phase synchronization, denoted by $S_p$, takes the form:
\begin{align}\label{14}
\mathrm{S}_{p} :=\frac{1}{2}\langle\delta p_{-}^{\prime 2}\rangle^{-1}\tag{9},
\end{align}
where $\delta p_{-}^{\prime}=\delta p_{1}^{\prime}-\delta p_{2}^{\prime},$ and $\delta \hat{p}_{j}^{\prime}$, the anti-Hermitian part of $\delta \hat{a}_{j}^{\prime}=[\delta \hat{q}_{j}^{\prime}+i \delta \hat{p}_{j}^{\prime}] / \sqrt{2}$, represents the quantum phase fluctuation operator of the subsystem $\mathrm{S}_{j}$, with $\delta \hat{a}_{j}^{\prime}=e^{-i \varphi_{j}} \delta \hat{a}_{j}$. $\hat{a}_{j}$ is the annihilation operator in the subsystem $S_j$.

In fact, the measure $\mathrm{S}_{p}$ is essentially the inverse of variance of the phase difference operators. Thus, it can be used to measure the level of the influence of quantum fluctuation on the classical phase synchronization, i.e., the greater $\mathrm{S}_{p}$ is, the smaller the variance is, which indicates the smaller influence of the quantum fluctuation on classical phase synchronization. Based on the uncertainty relation $\Delta \hat{q}^{\prime}_- \Delta \hat{p}^{\prime}_- \geq \hbar / 2$, one can deduce that the measure $\mathrm{S}_{p}$  can be arbitrary great with the help of squeezing resource, i.e., $\langle\delta p_{-}^{\prime 2}\rangle \rightarrow 0$ and $\langle\delta q_{-}^{\prime 2}\rangle \rightarrow \infty$ with $ q_{-}^{\prime}=\delta q_{1}^{\prime}-\delta q_{2}^{\prime}$ and $\delta \hat{q}_{j}^{\prime}$ being the Hermitian part of $\delta \hat{a}_{j}^{\prime}$. Thus, Ref.\cite{13} concluded that the phase synchronization is unbounded in quantum level.

Based on the definition of the phase anti-synchronization, the measure of phase anti-synchronization for the synchronized mode with same amplitude can be constructed as:
\begin{align}\label{15}
\mathrm{S}_{a} :=\frac{1}{2}\langle\delta p_{+}^{\prime 2}\rangle^{-1}\tag{10},
\end{align}
where $\delta p_{+}^{\prime}=\delta p_{1}^{\prime}+\delta p_{2}^{\prime}$. Similarly, one can deduce that the greater $\mathrm{S}_{a}$ is, the smaller the influence of the quantum fluctuation on classical phase anti-synchronization is. Also, one can deduce that, with the help of squeezing resource, the phase anti-synchronization is unbounded in quantum level, i.e., $\langle\delta p_{+}^{\prime 2}\rangle \rightarrow 0$ and $\langle\delta q_{+}^{\prime 2}\rangle \rightarrow \infty,$ with $ q_{+}^{\prime}=\delta q_{1}^{\prime}+\delta q_{2}^{\prime}$.

\section{Relationship between Phase anti-Synchronization and Quantum correlation} \label{C4}
Both synchronization and quantum correlation are associated with the correlations between two or more subsystems, and therefore, the research on the relationship between them is of great interest. Refs. \cite{13} and \cite{14} found that there exist some interest connection between quantum discord and synchronization. The phase anti-synchronization is also a kind of the synchronization, and thus the investigation of the relationship between quantum discord and phase anti-synchronization is necessary.

\begin{figure}
\centering 
\includegraphics[height=6cm]{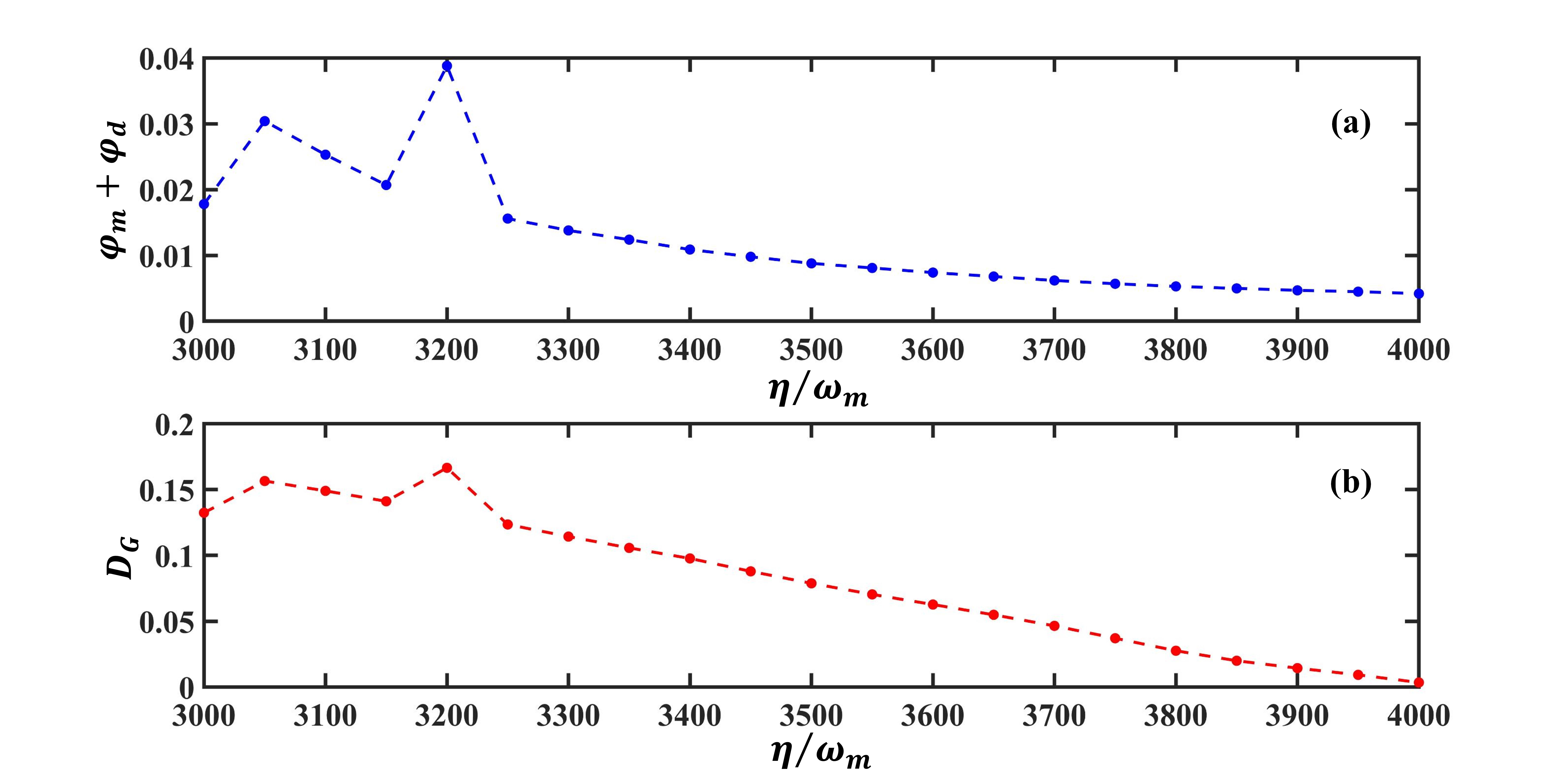}
\caption{Evolutions of $\varphi_{m}+\varphi_{d}$ (in (a)) and $D_G$ (in (b)) with respective to $\eta/\omega_m$.  Here $\varphi_{m}+\varphi_{d}$  and $D_G$ are the classical phase sum and quantum discord when phase anti-synchronization occurs, and the other related parameters are presented in the main text.} 
\label{f6}
\end{figure}

Ref. \cite{14} shows that the quantum discord has a correspondence to the classical phase synchronization jump at a certainty point. Here, we find that the relationship between quantum discord and classical phase anti-synchronization is different from the one between quantum discord and classical phase synchronization, i.e., quantum discord and classical phase anti-synchronization has the similar evolution with respect to $\eta$, as shown in Fig.\ref{f6}. Here, the quantum discord we used is quantified by the Gaussian quantum discord (for more detail, please see the Appendix C). Also, the relationship between entanglement and phase anti-synchronization has been investigated, and no positive result has been deduced, which is in agreement with Refs. \cite{13} and \cite{14}.

\section{Conclusion} \label{C5}
In conclusion, we have investigated the synchronization behavior between the mechanical oscillator and atomic ensemble within a common cavity, and find that sum of the phases of the two systems with different physical properties is locked after a long-time evolution. This phenomenon is referred to as the phase anti-synchronization, and this synchronization behavior is robust to the quantum fluctuation and thermal noise in the system under consideration. Such a synchronization behavior enriches our knowledge of the synchronization in the non-linear systems. Notably, the relationship between the phase anti-synchronization and quantum correlation has been investigated, and it is shown that there exist some positive connection between them.

\section*{Acknowledgments}
This work is supported by the National Natural Science Foundation of China (Grant No. 11574022) .

\section*{Appendix. A}
For the ensemble of $N$ effective two-level atoms, the total spin is $N/2$, and then one can introduce an effective atomic bosonic annihilation operator $\hat{d}$, which satisfies \cite{29}:
\begin{align}\label{a1}
\hat{S}_{z}&=\frac{N}{2}-\hat{d}^{\dagger} \hat{d},\nonumber \\
 \hat{S}_{+}&=\sqrt{N} \hat{d} \sqrt{\left(1-\hat{d}^{\dagger} \hat{d}\right)}, \nonumber \\
 \quad \hat{S}_{-}&=\sqrt{N} \hat{d}^{\dagger} \sqrt{\left(1-\hat{d}^{\dagger} \hat{d}\right)}\tag{A1}.
\end{align}
Under the condition that the atomic ensemble is fully inverted and $N$ is large, we have $\hat{d}^{\dagger} \hat{d} / N \ll 1$. Then, one can obtain that $\hat{S}_{+} \simeq \sqrt{N} \hat{d}$ and $\hat{S}_{-} \simeq \sqrt{N} \hat{d}^{\dagger}$, and the Hamiltonian (\ref{2}) can be rewritten as \cite{29,31}:
\begin{align}\label{a2}
\hat{H}_{a t}=- \hbar \omega_{\sigma} \hat{d}^{\dagger} \hat{d}+\hbar G \cos \left(\omega_{\mathrm{G}} t\right)\left(\hat{c}^{\dagger}+\hat{c}\right)\left(\hat{d}^{\dagger}+\hat{d}\right)\tag{A2},
\end{align}
 where $G=G_{0} \sqrt{N}$. Taking the resonance condition $\omega_G=\omega_L$, and applying the rotating wave approximation, one can obtain (in the interaction picture with respect to $\hbar \omega_{L} \hat{c}^{\dagger} \hat{c}$) \cite{29}:
\begin{align}\label{a3}
\hat{H}_{a t}=-\hbar \omega_{\sigma} \hat{d}^{\dagger} \hat{d}+\hbar \frac{G}{2}\left(\hat{c}^{\dagger}+\hat{c}\right)\left(\hat{d}^{\dagger}+\hat{d}\right)\tag{A3}.
\end{align}

\section*{Appendix. B}
The variance $\Delta\left(\hat{\varphi}_{m}+\hat{\varphi}_{d}\right)^{2}$ can be calculated by the covariance matrix $V(t)$, with the elements $V(t)[i, j]=[\langle u(i) u(j)+u(i) u(j)\rangle] / 2$ and $u=\left(\delta \hat{q}_{m}, \delta \hat{p}_{m}, \delta \hat{q}_{d}, \delta \hat{p}_{d}, \delta \hat{q}_{c}, \delta \hat{p}_{c}\right)^{T}$. Based on HL equation (\ref{8}), we can deduce that the time evolution of $V(t)$ is governed by \cite{37,41,42}:
\begin{align}\label{12}
\frac{d}{d t} V(t)=A V(t)+V(t) A^{T}+D\tag{B1},
\end{align}
where $D=\operatorname{diag}\{0, \gamma(2 \overline{n}+1), \Gamma, \Gamma, \kappa, \kappa\}$ is the diffusion matrix. The drift matrix $A$ reads:
\begin{align}\label{13}
A=\left(\begin{array}{cccccc}{0} & {\omega_{m}} & {0} & {0} & {0} & {0} \\ {-\omega_{m}} & {-\gamma} & {0} & {0} & {-g_{m} q_{c}} & {-g_{m} p_{c}} \\ {0} & {0} & {-\Gamma} & {-\omega_{\sigma}} & {0} & {0} \\ {0} & {0} & {\omega_{\sigma}} & {-\Gamma} & {-\sqrt{2} g_{m}} & {0} \\ {g_{m} p_{c}} & {0} & {0} & {0} & {-\kappa} & {g_{m} q_{m}+\Delta} \\ {-g_{m} q_{c}} & {0} & {-\sqrt{2} g_{m}} & {0} & {-\Delta-g_{m} q_{m}} & {-\kappa}\end{array}\right)\tag{B2}.
\end{align}
Assume that the mechanical oscillator is initially prepared in a thermal state corresponding to the temperature $T$, the atomic ensemble and the cavity mode fluctuations are initially in the vacuum state. Then, one can obtain the initial condition $V(0)=\operatorname{diag}[\overline{n}+1 / 2, \overline{n}+1 / 2,1 / 2,1 / 2,1 / 2,1 / 2]$. Taking advantage of Eq.(\ref{12}), one can obtain the numerical result of the variance $\Delta\left(\hat{\varphi}_{m}+\hat{\varphi}_{d}\right)^{2}$.
\section*{Appendix. C}
In fact, the covariance matrix $V(t)$ can be written as:
\begin{align}\label{c1}
V=\left[\begin{array}{ll}
{V_{A}} & {V_{C}} \\
{V_{C}^{T}} & {V_{B}}
\end{array}\right]\tag{C1}.
\end{align}
where $V_A$, $V_B$ and $V_C$ are $2\times2$ matrics, $V_A$ and $V_B$ account for the local properties of  modes $A$ and $B$, respectively, and $V_C$  describes the correlation between them. As we known, the Gaussian quantum discord is an asymmetric, and we here mainly focus on the Gaussian quantum A-discord, which is given by:
\begin{align}\label{c2}
D_G=f(\sqrt{\beta})-f\left(v_{-}\right)-f\left(v_{+}\right)-f(\sqrt{\varepsilon})\tag{C2}.
\end{align}
where $f(x)=\left(\frac{x+1}{2}\right) \log\left(\frac{x+1}{2}\right)-\left(\frac{x-1}{2}\right) \log\left(\frac{x-1}{2}\right)$, $v_{\pm}=\sqrt{\frac{\Sigma_{+} \pm \sqrt{\Sigma_{+}^{2}-4 \operatorname{det} V}}{2}}$, and $\Sigma_{\pm}=\operatorname{det} V_{A}+\operatorname{det} V_{B} \pm 2 \operatorname{det} V_{C}$. $\varepsilon$ is given by;
\begin{align}\label{c3}
\varepsilon=\left\{\begin{array}{lr}
{\frac{2 \gamma^{2}+(\beta-1)(\delta-\alpha)+2|\gamma| \sqrt{\gamma^{2}+(\beta-1)(\delta-\alpha)}}{(\beta-1)^{2}},} & {\frac{(\delta-\alpha \beta)^{2}}{(\beta+1) \gamma^{2}(\alpha+\delta)} \leqslant 1} \\
{\frac{\alpha \beta-\gamma^{2}+\delta-\sqrt{\gamma^{4}+(\delta-\alpha \beta)^{2}-2 \gamma^{2}(\delta+\alpha \beta)}}{2 \beta},} & {\text { otherwise }}
\end{array}\right.\tag{C3}.
\end{align}
where $\alpha=\operatorname{det} V_{A}$, $\beta=\operatorname{det} V_{B}$, $\gamma=\operatorname{det} V_{C}$,  and $\delta=\operatorname{det} V$. Then, using Eqs.(\ref{12}) and (\ref{c2}), one can obtain the  Gaussian quantum discord between the mechanical oscillator and the atomic ensemble.


\begin{thebibliography}{99}
\bibitem{1}J. A. Acebrón, L. L. Bonilla, C. J. Pérez Vicente, F. Ritort, R. Spigler, Rev. Mod. Phys. \textbf{77}, 137 (2005).
\bibitem{2}M. Kapitaniak, K. Czolczynski, P. Perlikowski, A. Stefanski, T. Kapitaniak, Phys. Rep. \textbf{517}, 1 (2012).
\bibitem{3}M. Maianti, S. Pagliara, G. Galimberti, F. Parmigiani, Am. J. Phys. \textbf{77}, 834 (2009).
\bibitem{4}J. Pantaleone, Am. J. Phys. \textbf{70}, 992 (2002).
\bibitem{5}M. Aguiar, P. Ashwin, A. Dias, and M. Field, J. Nonlin. Sci. \textbf{21}, 271 (2011).
\bibitem{6}B. Blasius, A. Huppert, and L. Stone, Nature \textbf{399}, 354 (1999).
\bibitem{7}N. Porat-Shliom, Y. Chen, M. Tora, A. Shitara, A. Masedunskas, and R. Weigert, Cell Reports \textbf{9}, 514 (2014).
\bibitem{8}L. M. Pecora and T. L. Carroll, Phys. Rev. Lett. \textbf{64}, 821 (1990)
\bibitem{9}S. H. Strogatz, I. Stewart, Sci. Am. \textbf{269}, 102 (1993).
\bibitem{10}L. Angelini, G. Lattanzi, R. Maestri, D. Marinazzo, G. Nardulli, L. Nitti, M. Pellicoro, G. D. Pinna, S. Stramaglia, Phys. Rev. E \textbf{69}, 061923 (2004).
\bibitem{11}A. Arenas, A. Diaz-Guilera, J. Kurths, Y. Moreno, C. Zhou, Phys. Rep. \textbf{469}, 93 (2008).
\bibitem{12}E. Padmanaban, S. Boccaletti, S. K. Dana, Phys. Rev. E \textbf{91}, 022920 (2015).
\bibitem{13}A. Mari, A. Farace, N. Didier, V. Giovannetti, R. Fazio, Phys. Rev. Lett. \textbf{111}, 103605 (2013).
\bibitem{14}F. Bemani, A. Motazedifard, R. Roknizadeh, M. H. Naderi, D. Vitali, Phys. Rev. A \textbf{96}, 023805 (2017).
\bibitem{15}B. Militello, H. Nakazato, A. Napoli, Phys. Rev. A \textbf{96}, 023862 (2017).
\bibitem{16}L. Ying, Y. C. Lai, and C. Grebogi Phys. Rev. A \textbf{90}, 053810 (2014).
\bibitem{17}M. Xu, D. A. Tieri, E. C. Fine, J. K. Thompson, M. J. Holland, Phys. Rev. Lett. \textbf{113}, 154101 (2014).
\bibitem{18}C. G. Liao, R. X. Chen, H. Xie, M. Y. He, and X. M. Lin, Phys. Rev. A \textbf{99}, 033818 (2019).
\bibitem{19}C. Davis-Tilley, C. K. Teoh and A. D. Armour, New J. Phys. \textbf{20}, 113002 (2018).
\bibitem{20}V. Ameri, M. Eghbali-Arani, A. Mari, A. Farace, F. Kheirandish, V. Giovannetti, and R. Fazio, Phys. Rev. A \textbf{91}, 012301 (2015).
\bibitem{21}M. R. Hush, W. B. Li, S. Genway, I. Lesanovsky, and A. D. Armour, Phys. Rev. A \textbf{91}, 061401(R) (2015).
\bibitem{22}A. Roulet, and C. Bruder, Phys. Rev. Lett. \textbf{121}, 063601 (2018).
\bibitem{23}C. Davis-Tilley and A. D. Armour, A \textbf{94}, 063819 (2016).
\bibitem{24}E. Schr\"{o}dinger, Physikalisch-mathematische Klasse \textbf{14}, 296 (1930).
\bibitem{25}V. V. Dodonov, Phys. Rev. A \textbf{97}, 022105 (2018).
\bibitem{26}M. Xu and M. J. Holland, Phys. Rev. Lett. \textbf{114}, 103601 (2015).
\bibitem{27}H. Qiu, R. Zambrini, A. Polls, J. Martorell, and B. Julia-Diaz, Phys. Rev. A \textbf{92}, 043619 (2015).
\bibitem{28}M. Aspelmeyer, T. J. Kippenberg, F. Marquardt, Rev. Mod. Phys. \textbf{86}, 1391 (2014).
\bibitem{29}A. Motazedifard, F. Bemani, M. H. Naderi, R. Roknizadeh, and D Vitali, New J. Phys. \textbf{18}, 073040 (2016).
\bibitem{30}M. H. Wimmer, D. Steinmeyer, K.Hammerer, and M. Heurs, Phys. Rev. A \textbf{89}, 053836 (2014).
\bibitem{31}F. Bariani, H. Seok, S. Singh, M. Vengalattore, and P. Meystre, Phys. Rev. A \textbf{92}, 043817 (2015).

\bibitem{3a1}F. Bemani, R. Roknizadeh, A. Motazedifard, M. H. Naderi, and D. Vitali, Phys. Rev. A \textbf{99}, 063814 (2019).
\bibitem{3a2}B. D. Hauer, T. J. Clark, P. H. Kim, C. Doolin, and J. P. Davis, Phys. Rev. A \textbf{99}, 053803 (2019).
\bibitem{3a3}G. F. Huang, W. W. Deng, H. T. Tan, and G. L. Cheng,  Phys. Rev. A \textbf{99}, 043819 (2019).

\bibitem{32}K. Hammerer, A. S. Sorensen, and E. S. Polzik, Rev. Mod. Phys. \textbf{82}, 1041 (2010).
\bibitem{33}V. Giovannetti, and D. Vitali, Phys. Rev. A \textbf{63}, 023812 (2001).
\bibitem{34}R. Benguria, M. Kac, Phys. Rev. Lett. \textbf{46}, 1 (1981).
\bibitem{35}T. E. Lee and H. R. Sadeghpour, Phys. Rev. Lett. 111, 234101 (2013).
\bibitem{36}A. Dantan, C. Genes, D. Vitali, M. Pinard, Phys. Rev. A \textbf{77}, 011804(R) (2008).
\bibitem{37}A. Mari, J. Eisert, Phys. Rev. Lett. \textbf{103}, 213603 (2009).
\bibitem{38}W. K. Wootters, Phys. Rev. Lett. \textbf{80}, 2245 (1998).
\bibitem{39}C. W. Gardiner and P. Zoller 2000 Quantum Noise (Berlin: Springer).
\bibitem{40}C. Genes, D. Vitali, and P. Tombesi, Phys. Rev. A \textbf{77}, 050307(R) (2008)
\bibitem{41}S. Walter, A. Nunnenkamp, and C. Bruder, Phys. Rev. Lett. \textbf{112}, 094102 (2014).
\bibitem{42}M. Ludwig and F. Marquardt, Phys. Rev. Lett. \textbf{111}, 073603 (2013).


\end{thebibliography}
\end{document}